\documentclass{iau} 
\usepackage{graphicx}

\newcommand{\hi}{H{\sc i}}
\newcommand{\hii}{H{\sc ii}}
\newcommand{\hdue}{H$_2$}
\newcommand{\msun}{M$_{\odot}$}

\title[Evolution of star-forming dwarf galaxies in different environments] 
{Evolution of star-forming dwarf galaxies in different environments}

\author[Marco Grossi]   
{Marco Grossi$^1$}

\affiliation{$^1$Observat\'orio do Valongo, Universidade Federal do Rio de Janeiro, \\ Ladeira Pedro Ant\^onio 43,
Rio de Janeiro  20080-090, Brazil\\ email: {\tt grossi@astro.ufrj.br} 
}

\pubyear{2018}
\volume{S344}  
\jname{Dwarf Galaxies: From the Deep Universe to the Present}
\editors{K. McQuinn, \& S. Stierwalt, eds.}
\begin{document}

\maketitle

\begin{abstract}
The ubiquity of star-forming dwarf galaxies (SFDG) in the local Universe allows us to trace their evolution in all type of environments,
from voids to rich clusters.
SFDGs in low-density regions are still assembling their mass, they often show peculiar gas morphology and kinematics, likely associated
to external gas accretion or galaxy interactions, and they can experience strong bursts of star formation.
The most metal-poor SFDGs are found in the field and they are unique laboratories
to investigate the star formation process in the low-metallicity regime, at conditions similar to their high-redshift analogues.
On the other hand, SFDGs in intermediate- and high-density environments provide a
key to understand the processes that remove their interstellar medium (ISM) and suppress star formation, 
leading to the different types of gas-poor early-type dwarfs.
We review the most recent results on the properties of SFDGs at low and high galaxy densities focusing in
particular on the impact of a cluster environment on their ISM components (dust, molecular,
atomic and ionised gas). We analyse the population of SFDGs in the nearest rich
clusters: Virgo, which is still in the process of assembly, and Fornax, which is more dynamically
evolved, more compact and denser. We discuss how the different evolutionary stage of the two structures
affects the properties of SFDGs.
\end{abstract}

\firstsection 
\section{Introduction}


More than 70\% of all galaxies in the Local Volume are star-forming dwarf galaxies (SFDGs; Karachentsev et al. 2004), i.e. 
low-mass systems rich in atomic gas (\hi) and a low metal
and dust content. The properties of their interstellar medium (ISM) resemble those of the fundamental building
block of galaxies, thus they can be considered
{\em ``the initial state of all dwarf galaxies``} (Bergvall 2011).
Because they are ubiquitous in the Local Universe they can be studied in different environments,
from voids to rich clusters. 
At low galaxy densities they allow us to study the process of mass assembly unaffected by the interplay with massive galaxies occurring
in clusters and groups. 
At high galaxy densities they can provide a key to understand 
how different types of dwarf galaxies can emerge due to the interaction
with their surroundings. 

The $\Lambda$ cold dark matter ($\Lambda$CDM) model (Weinberg et al. 2004) predicts that galaxies can  grow in two ways: 
they assemble hierarchically through mergers of smaller dark matter halos, or 
they accrete gas from the cosmic web.
Low-mass systems constitute the dominant population
at all redshifts and they play a crucial role in the hierarchical build-up of galaxies.
According to simulations dwarf galaxies 
experience on average three major mergers in their lifetime
(Fakhouri et al. 2010).
Dwarf-dwarf mergers are predicted to be more common at earlier times (Klimentowski et al. 2010), 
however
dwarf galaxies in isolated environments are expected to be twice more likely to experience
a recent merger at $z < 1$ compared to satellites of a massive host (Deason et al. 2014).
Therefore, assessing the role of mergers and interactions between dwarfs is extremely important to understand their impact
on the evolution of these systems.
While massive galaxy mergers have been studied in great detail, dwarf-dwarf interactions have yet to be 
thoroughly explored, also because such systems are more difficult to be observed. 

Diffuse gas accretion is another important process affecting
the process of mass assembly in galaxies (Kere{\v s} et al. 2005).
Models of galaxy formation predict that gas accretion from the cosmic web is a primary driver of star formation
over cosmic history (Dekel et al. 2013), and that
at halo masses of M$_{h} = 10^9 - 10^{10}$ M$_{\odot}$ cold gas accretion is still expected to occur
at $z = 0$ in low-density environments (Kere{\v s} et al. 2009).
Low-mass galaxies should accrete most of their gas
through cold flows reaching the central parts of the dark matter halo
without being shock heated to the virial temperature (Dekel \&
Birnboim 2006). 

Another issue is the role played by environment on galaxy
evolution. 
Indeed one of the most fundamental
correlations between the properties of galaxies in the local Universe is the so-called 
morphology-density relation (Dressler 1980, Dressler et al. 1997). 
Several studies
have shown that early-type, quiescent galaxies are preferentially found in denser environments
and such a correlation is observed out to z $\sim$ 1 (Kauffmann
et al. 2004; 
Scoville et al. 2013; Darvish et al. 2016).
Environmental effects on galaxy evolution are expected
to be stronger in dwarfs.
Because of their lower gravitational
potentials and less dense ISM (Bolatto et al. 2008),
SFDGs are more sensitive to their surroundings than more massive galaxies 
(Boselli \& Gavazzi 2006). 
Ram-pressure stripping (Gunn \& Gott 1972), starvation (Larson et al. 1980), tidal interactions (Brosch et al. 2004), 
and galaxy harassment (Moore et al. 1996) are among the suggested processes that can modify the morphology of a dwarf galaxy in clusters and 
quench the star formation process through the removal
of their ISM, 
However, the transformation path from star-forming
to passive dwarfs is still poorly understood, as well as the nature of the late-type progenitors of today
dwarf ellipticals (dEs; Lisker et al. 2007, 2013; Mistani et al. 2016).

Here we briefly review the most recent results on the properties of SFDGs 
focusing in particular on two types of environment, the very low density (such as voids) and the very high densities (i.e. the nearest 
rich clusters to us, Virgo and Fornax), and we discuss
how studying SFDGs in the nearby Universe 
can help clarifying our understanding of 
galaxy evolution at the low-mass regime.

\section{Low-density environments}

\subsection{Very isolated star-forming dwarfs: evidence of gas accretion}

Cosmic voids are vast underdense regions with sizes of
20 - 50$h^{-1}$ Mpc and approximately spherical shape 
 (Einasto et al. 1980, El-Ad \& Piran 1997, Tully 2008).
Largely unaffected by the processes modifying galaxies
in clusters and groups, voids represent an extreme and pristine environment
in which to study the accretion history and the process of mass assembly in galaxies.

The advent of the 2 degree field Galaxy Redshift Survey (2dFGRS; Colless et al. 2001) and of the  Sloan Digital Sky Survey (SDSS; York
et al. 2000) provided the first large samples 
of void galaxies that could be used to  
assess their statistical properties 
(Hoyle \& Vogeley 2004; Pan et al. 2012).
Observations 
showed that galaxies in underdense environments are fainter, bluer, and less massive than those in denser regions 
(Hoyle et al. 2005; Moorman et al. 2015), 
and that the centre of voids are mostly populated by dwarf systems
 (Hoyle et al. 2012).

\begin{figure}[t]
\begin{center}
\includegraphics[bb=515 0 0 720,angle = -90, width=7.3cm, clip]{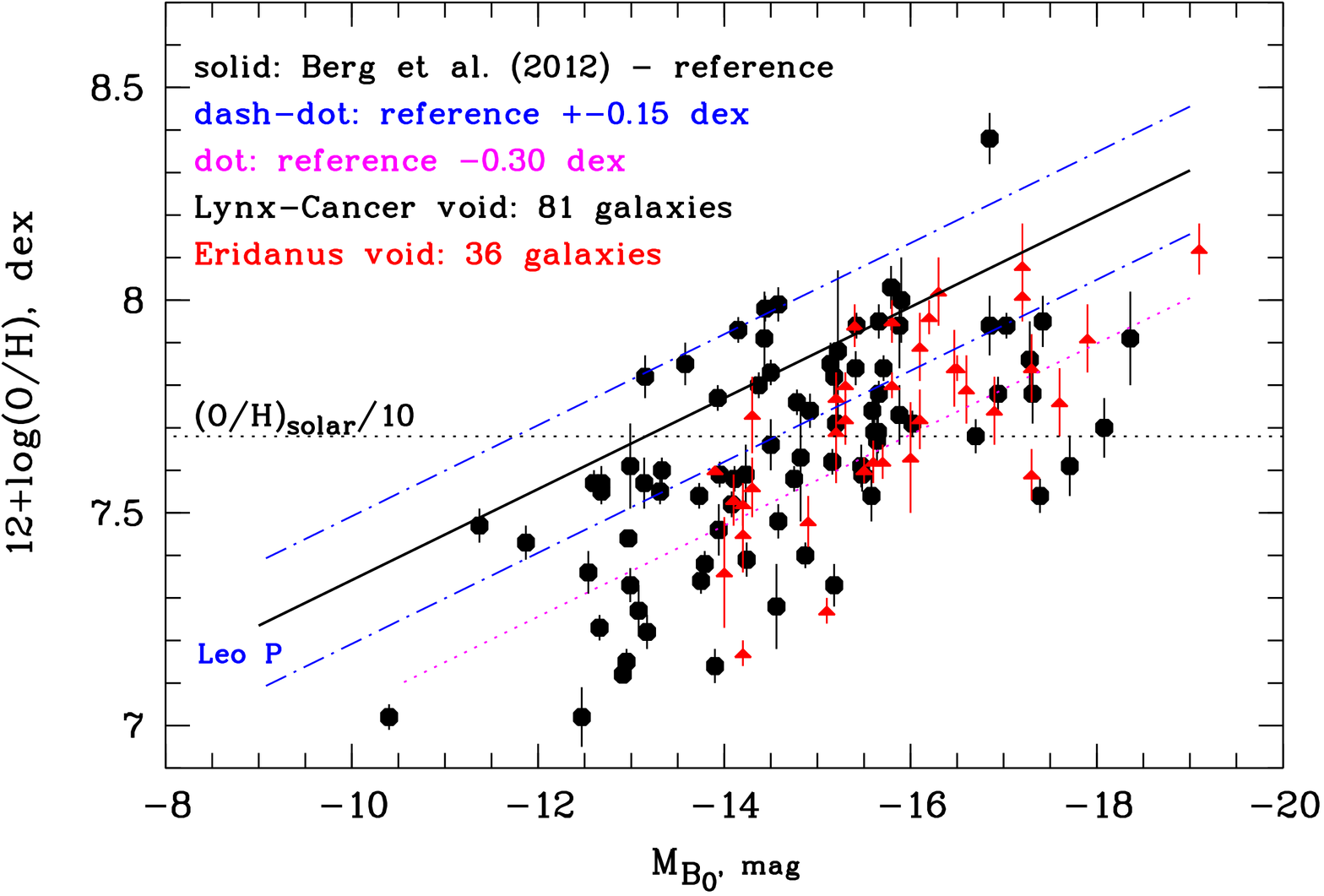}
\includegraphics[bb= 50 360 750 990,width=6cm,clip]{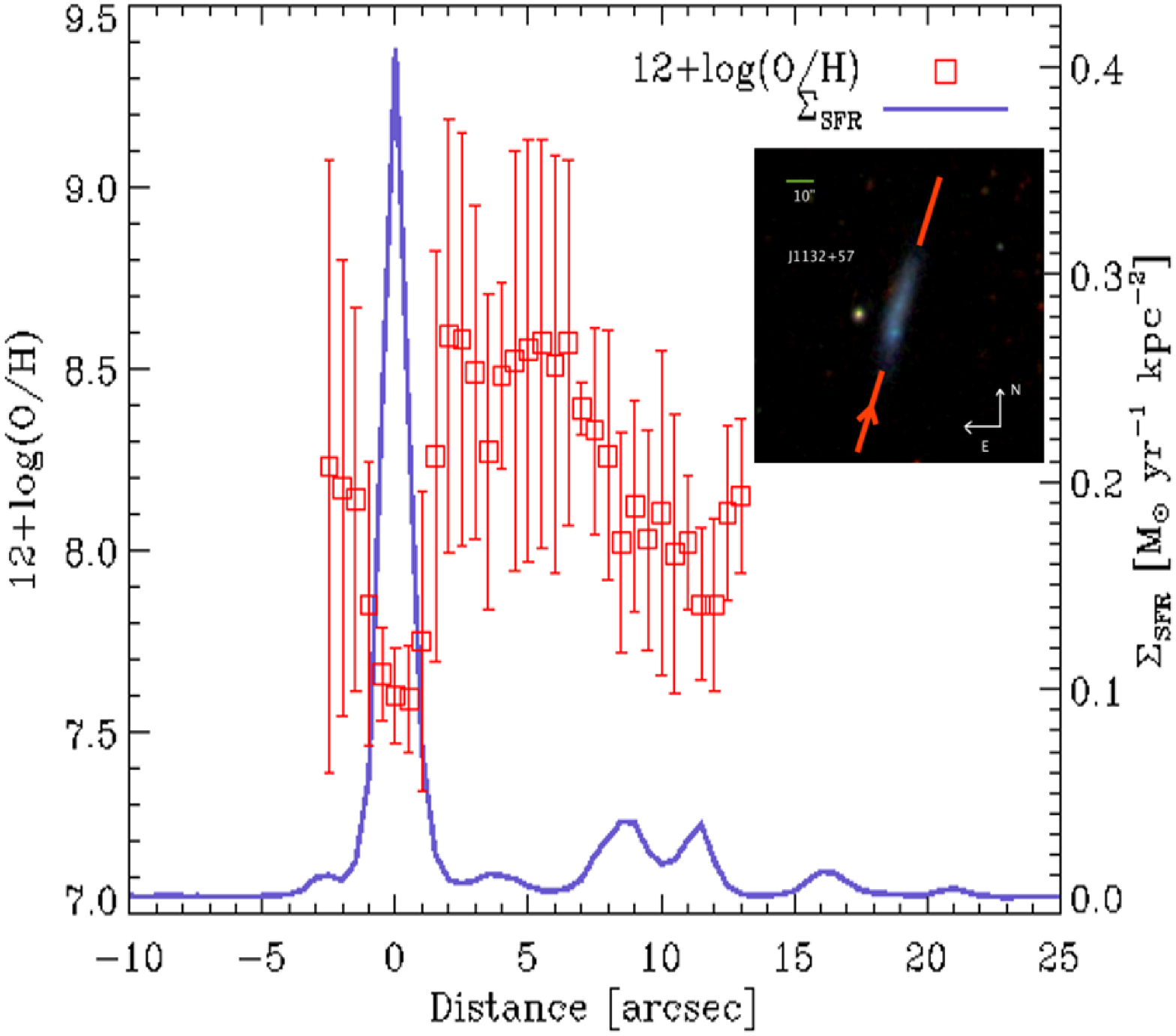}
\caption{Left: Luminosity-metallicity relation for galaxies in the Eridanus (triangles) and Lynx-Cancer (filled-dots)
voids. 
The solid line shows the linear regression obtained for the control sample of galaxies in the Local Volume (Berg et al. 2012).
The dotted line (offset by $-$0.30 dex from the reference sample)
indicates the region where the most metal-poor dwarfs are located.
Adapted from Kniezev et al. (2018). Right: Variation of oxygen abundance
(squares with error bars) and surface SFR (solid line) along the major
axis of the XMP galaxy SDSS~J1132+57. The metallicity drop coincides with the brightest knot and the peak 
of the surface SFR. The SDSS cut-out image of the galaxy is shown in the upper-right corner of the panel. Adapted from S\'anchez-Almeida et al. 2015.}
\label{fig1}
\end{center}
\end{figure}

At magnitudes brighter than $M_{B,r} \sim -16$, probed 
in most of the surveys of rather distant voids extracted from the SDSS (D $\sim$ 100-200 Mpc),
there is not a clear difference between 
SFDGs in very isolated environments or in denser regions 
in terms of specific star formation rates (sSFR) and gas fractions (Kreckel et al. 2011a, 2012; Moorman et al. 2016).
On the other hand, studies of the nearest voids such as the Lynx-Cancer or Eridanus, 
allowed to inspect the
faintest isolated dwarfs ($M_r > -14$ mag, Pustilnik 2011a; Kniazev 2018), 
revealing a population of
extremely gas-rich, very metal-poor objects (Z $<$ Z$_{\odot}$/20), with 
low-surface brightness (LSB) optical 
counterparts (Pustilnik et al. 2011b, 2013).
They show metallicities lower by
at least 0.2 dex compared to galaxies of similar luminosity.
The faintest objects have oxygen abundances 2 to 4 times lower than the galaxies in the Local Volume 
(Fig \ref{fig1}, left panel; Pustilnik et al. 2016; Kniazev et al. 2018).
This suggests that low-luminosity SFDGs in voids evolve slower, 
likely due to the much reduced rate of galaxy interactions, and that the main star formation episode
 occurred rather recently (between one and a few
Gyrs ago; Pustilnik et al. 2013).

Recent surveys have found several unusual
objects in voids. 
VGS12 is an isolated low-mass system showing a \hi\ polar disc with no optical counterpart;
the absence of a companion makes it unlikely
that the substantial neutral hydrogen content (M$_{HI} = 3 \times 10^9$ \msun, $f_{gas} \sim 0.8$) is the result of a close encounter
(Stanonik et al. 2009).
KK246 is another very isolated dwarf galaxy in the Tully void; the extended \hi\ disc shows
anomalous kinematics associated to 
a gas cloud whose velocity is inconsistent with the disc rotation
(Kreckel et al. 2011b).  
These works suggest that 
the presence of extended and disturbed \hi\ morphologies around these objects 
may arise from material flow inside 
the voids and that SFDGs in underdense environments are still assembling their mass through gas accretion from the
intergalactic medium (IGM). 

Another evidence of ongoing intergalactic gas accretion in SFDGs is provided by 
metallicity measurements of their \hii\ regions.
Extremely metal-poor star-forming dwarfs (XMPs), with oxygen abundances 12 + log O/H $\lesssim$ 7.7 ($< 0.1$ Z$_{\odot}$),
constitute a rare 
class of isolated galaxies in the nearby Universe
(S\'anchez-Almeida et al. 2016).
They are considered 
to be the best local analogues of the population of 
dwarf galaxies at high redshifts.
XMPs are predominantly low-mass systems, with large amounts of \hi\ (M$_{HI}$/M$_* \sim$ 20), and
 they do not follow the mass-metallicity relation (S\'anchez-Almeida et
al. 2016). 
Usually they show an off-center star-forming region of lower
metallicity compared to the rest of the disc (Fig. \ref{fig1}, right panel). 
The decrease of the metal abundance in these regions is found to be of the order of 0.3 dex or
larger and it is associated to the peak of the 
surface star formation rate (SFR; S\'anchez-Almeida et al. 2015).
The metallicity drops 
may give evidence of pristine gas accretion from the IGM.  
The infall of metal-poor gas triggers star formation and mixes with the more
metal-rich, pre-existing gas component (S\'anchez-Almeida et al. 2016).
If the gas accretion and the starburst occur
in a time scale similar to or shorter than the mixing
time scale, the mass of the star-forming clumps is expected to
be dominated by the metal-poor gas component  (van de Voort \& Schaye 2012).

One of the most intriguing properties of XMPs is the
existence of a lower metallicity threshold detected in their star-forming regions, usually of the order of
2\% the solar metallicity. If the star formation activity is fed by the infall of gas from the IGM, the 
observed minimum metallicity could be related to the current abundance of the local intergalactic gas.
Indeed simulations predict that the metal abundance of the IGM has been raising with time, 
enriched by galactic winds and outflows, and its current value is expected to be of the order of 1\% of
the solar value (van de Voort \& Schaye 2012, Rahmati et al. 2016).

\subsection{Dwarf-dwarf interactions}

Mergers of massive gas-rich galaxies are associated with 
bursts of star formation and they play an important role in the growth
and evolution of galaxies 
(Mihos \& Hernquist 1994; Bournaud et al. 2011).
However, the mechanisms that trigger the enhancement of the star-formation activity
in low-mass systems are poorly understood.
Mergers and interactions between dwarf galaxies have been
proposed to explain the properties of starbursting blue compact dwarfs (BCDs;
 Noeske et al. 2001, Bekki 2008) and of  
actively star-forming dwarfs without no clear optical companion (Telles \& Terlevich 1995).
Several SFDGs show extended
and filamentary \hi\ structures, which may indicate a
recent interaction or a merger event (Stil \& Israel 2002; Lelli et al. 2014).
High resolution 21-cm observations of IZw18, the prototypical BCD, unvealed large amounts of \hi\ around the
galaxy 
with a spectacular \hi\ plume extending for over 13.5 kpc, probably originated in a tidal interaction between
the main complex (IZw18A) and the lower-mass companion IZw18C (Lelli et al. 2012).

\begin{figure}[t]
\begin{center}
\includegraphics[bb=30 22 290 246,width=6.3cm, clip]{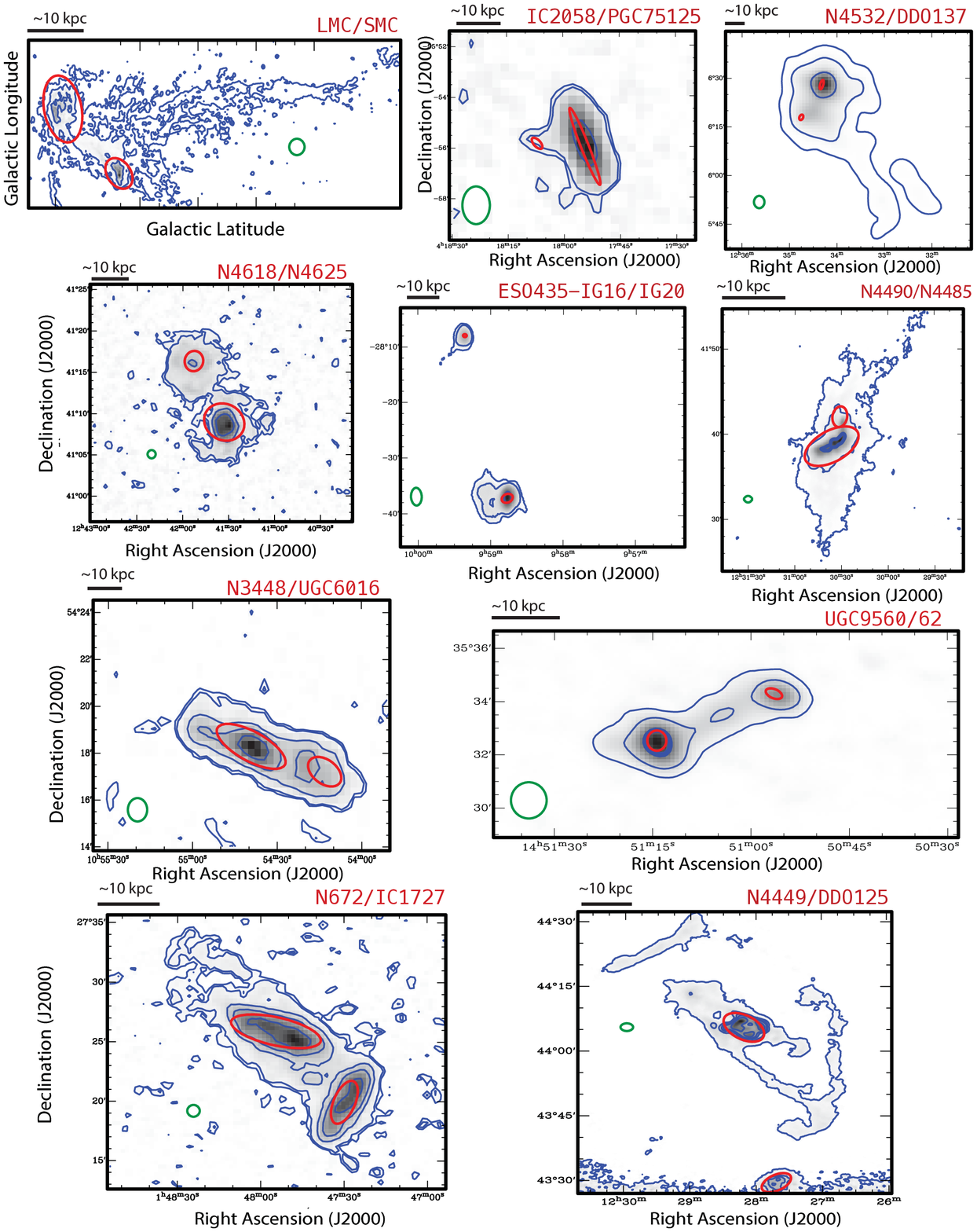}
\includegraphics[bb=330 25 570 246,width=5.8cm, clip]{fig02.eps}
\includegraphics[bb=30 445 235 630,width=6.6cm, clip]{fig02.eps}
\includegraphics[bb=432 415 610 620,width=5.2cm, clip]{fig02.eps}
\caption{\hi\ maps of a subset of dwarf galaxy pairs taken from the LV-TNT survey. The ellipses indicate the 
size of the stellar disc of each galaxy. Adapted from Pearson et al. (2016).}
\label{fig2}
\end{center}
\end{figure}

The first stellar stream discovered around a dwarf irregular galaxy --
NGC~4449, one of the most intensely star-forming systems in the Local Volume --
showed that satellite accretion occurs also in SFDGs (Mart\'inez-Delgado et al. 2012).
The stream is associated to the tidal disruption of a dwarf spheroidal -- NGC~4449B, with a stellar mass of at least 1/50 of the primary galaxy --
and the merger is probably responsible for the enhanced star formation activity of NGC~4449.
Examples of mergers/interactions of dwarf galaxies are also being found in voids.
The very metal-poor SFDG DDO~68 (Ekta et al. 2008) shows extended stellar features (a tail and an arc) 
that give evidence  of a multiple
merging event (Annibali et al. 2016), 
while a much fainter companion, DDO~68C
at $\sim$42 kpc, is connected to the primary galaxy by a LSB \hi\ bridge (Cannon et al. 2014).

Discoveries of dwarf galaxy pairs are increasing (Stierwalt et al. 2015, Paudel et al. 2018). 
The TiNy Titans (TNT) Survey (Stierwalt et al. 2015) provided a first systematic study
of 104 dwarf galaxy pairs in different interaction stages
and environments, selected from the SDSS. The survey, which includes pairs with  
masses between 7 $<$ log(M$_∗$/M$_{\odot}$) $<$ 9.7, mass ratios of M$_{*,1}$/M$_{*,2}$ $<$ 10 and a projected separation smaller
than 50 kpc, showed that 
interactions between low mass galaxies do affect the
structure and star formation activity of SFDGs. The SFR increases in more compact isolated pairs, similarly
to what found in massive galaxies (Patton et al. 2013), confirming the role of close encounters in enhancing star formation activity in low-mass
systems.

\hi\ maps of dwarf galaxy pairs, 
(the Local Volume TiNy Titans survey, LV-TNT; Pearson et al. 2016)
pointed out that large amounts of
gas are present in their outskirts compared to non-paired analogues (more than 50 per cent of the total gas mass is beyond
their stellar extents). 
This suggests that dwarf-dwarf interactions can move gas out to large distances from the pair 
(Fig. \ref{fig2}). 
However, simulations of the dwarf system NGC~4490 - NGC~4485, an isolated pair 
surrounded by a \hi\ envelope of $\sim$ 50 kpc size (Pearson et al. 2018; Fig. \ref{fig2}, bottom-right panel), 
predict that the gas structure remains bound to the system after the interaction and it will 
be re-accreted in a few Gyrs, providing a long-lasting gas supply for future star formation episodes (Pearson et al. 2018).
On the other hand, if the pairs are non-isolated, the interaction with the hot halo of the nearby massive galaxy will prevent gas from
being re-accreted, as in the
case of the Magellanic Clouds (Pearson et al. 2016). 

Close encounters among gas-rich dwarf galaxies can have a considerable impact on the star formation
history and dynamical state of these systems, however they are rare in the local Universe.
According to Stierwalt et al. (2017), within the completeness limits of the SDSS, only $\sim$4\% of dwarf galaxies at $z < 0.015$ have close companions (i.e. a separation 
smaller than 50 kpc) with a stellar mass
ratio of at least 1:10.

\begin{figure}[t]
\begin{center}
\includegraphics[bb=30 60 2000 660,width=14cm, clip]{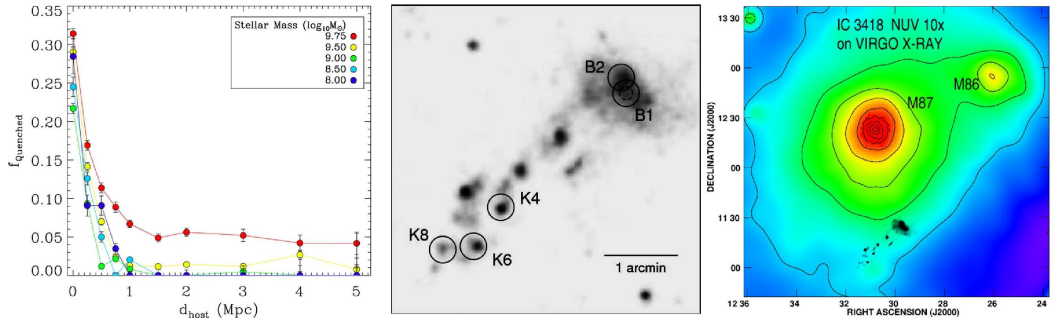}
\caption{Left panel: Fraction of quenched dwarf galaxies as a function of distance to the nearest massive host in different stellar
mass bins.  Adapted from Geha et al. (2012). Central panel: GALEX image of the dwarf irregular galaxy IC~3418 showing the main stellar body and 
the star-forming regions in the extended tail. Adapted from J\'achym et al. (2013).
Right panel: Location of IC~3418 with respect to M87, the massive elliptical at the
core of the Virgo cluster. Contours represent X-ray emission measured by ROSAT. The size of IC~3418 is increased by a factor 10. 
Adapted from Kenney et al. (2014).}
\label{fig3}
\end{center}
\end{figure}

\section{High density environments}

Morphological segregation is
the clearest evidence of the processes that determine the evolution of galaxies in high-density 
environments and it 
applies to both high- and low-mass systems.
Early-type dwarfs (dEs and dwarf spheroidals, dSphs) 
represent  the dominant population
in clusters, but they are very rarely found in the field.
According to Geha et al. (2012) the fraction of quenched dwarfs  
in low-density environments\footnote{The study analysed dwarf galaxies in the mass range 10$^7 <$ M$_* < 10^9$ M$_{\odot}$} 
(i.e. beyond 1.5 Mpc from a massive host) is less 
than 0.06\%.
On the other hand, they represent $\sim$ 30\% of the dwarf galaxy population in dense environments, i.e. when they are found at a distance of 250 kpc or less from 
a massive host (Fig. \ref{fig3}, left panel).

It is well established that late-type galaxies in rich clusters tend to have a
lower \hi\ content than their more isolated counterparts and that there is an
anti-correlation between \hi\ deficiency and the distance to the cluster center 
(Giovanardi et al. 1983; Haynes et al. 1984; Chung et al. 2009; Denes et al. 2014). 
\hi\ removal is the first step to quench the SF activity, and it usually
corresponds to a truncation of the H$\alpha$ emission in the outer
regions of galaxy discs (Koopman \& Kenney 2004; Boselli \& Gavazzi 2006).

It was long debated whether this also applies to the other components of the ISM such as molecular gas,
which is generally more concentrated in the center of galaxies than \hi\ (Kenney \& Young
1986; Fumagalli et al. 2009) and dust (Popescu et al. 2002; Tuffs et al. 2002). 
Spiral galaxies in the Virgo cluster show that lower-density atomic gas is preferentially removed from the outskirts and that
\hi-deficient spirals have dust discs significantly less extended than gas-rich ones (Cortese et al. 2010, 2012; Corbelli et al. 2012).
On the other hand, the molecular gas component appears to be removed less efficiently than 
\hi\ in cluster galaxies (Boselli et al. 2014, Chung et al. 2017).

An example of the dramatic effects of a dense environment on the transformation of SFDGs is given by 
IC~3418, a \hi- and \hdue-deficient dwarf in the Virgo cluster (J\'achym
et al. 2013). The galaxy is at 
$\sim 280$ kpc in projection from the massive elliptical M87 at the core of the cluster,
with less than 1\% of the expected \hi\ and \hdue\ mass. All the 
atomic gas has been stripped from its
main body, and CO is only marginally detected in the center. 
The star formation is quenched in the disc of this galaxy, 
but it shows a tail of star-forming regions originated from ram-pressure stripped gas
(Fig. \ref{fig3}, central and right panels; J\'achym
et al. 2013; Kenney et al. 2014). 

\begin{figure}[t]
\begin{center}
\includegraphics[bb=60  15 540 550,width=6.7cm]{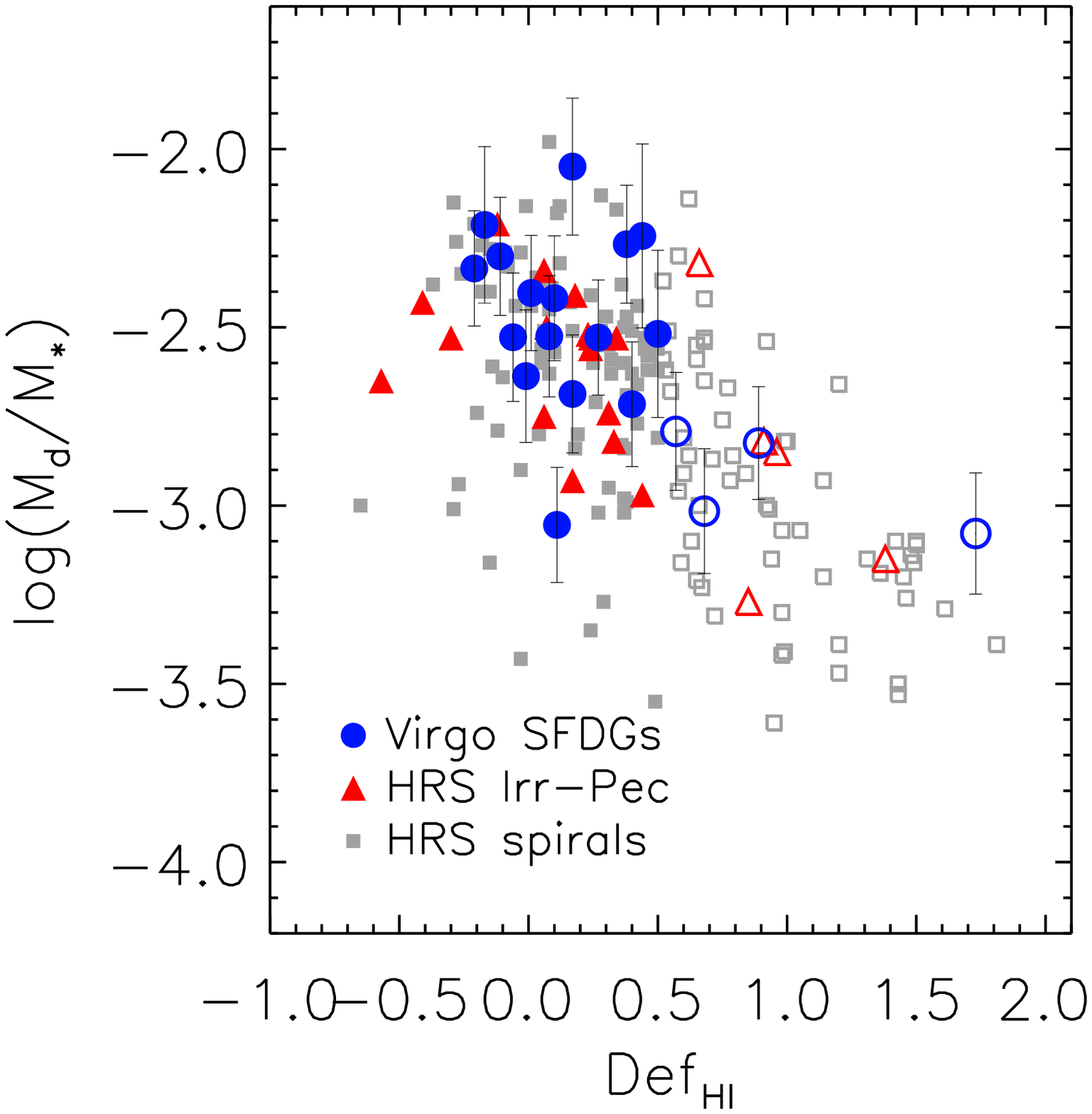}
\includegraphics[bb=60  15 540 550,width=6.7cm]{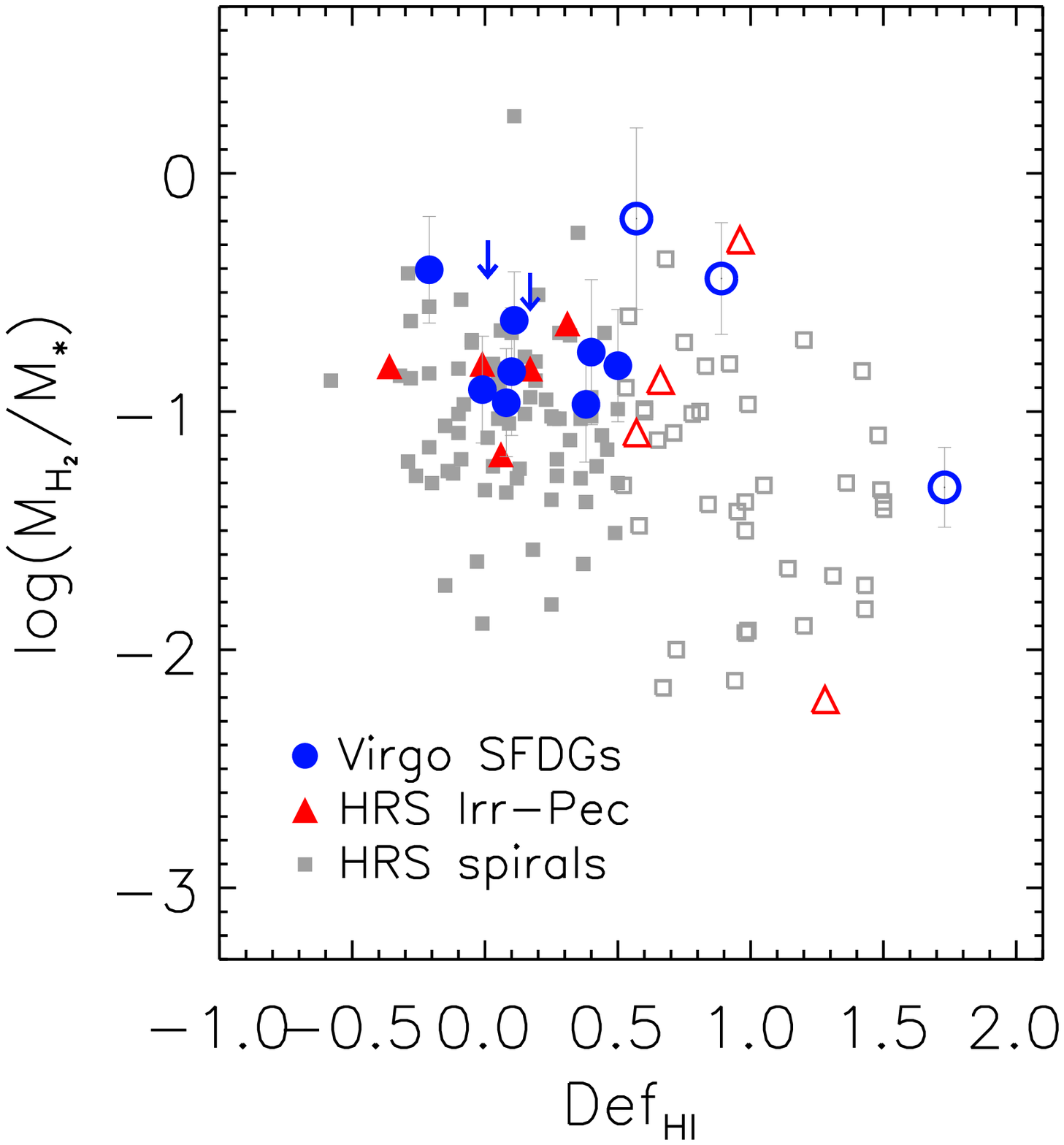}
\caption{Environmental effects on the different components of the interstellar medium of Virgo SFDGs (dots) compared to
the Herschel Reference Survey spiral (squares) and irregular-peculiar (triangles) galaxies. H{\sc i}-deficient galaxies are indicated with
empty symbols, and arrows indicate Virgo dwarf upper limits.
Left: Ratio of dust-to-stellar mass against \hi\ deficiency. Right: Ratio of \hdue-to-stellar mass against \hi\ deficiency. Adapted
from Grossi et al. (2016).} 
\label{fig4}
\end{center}
\end{figure}

\begin{figure}[t]
\begin{center}
\includegraphics[bb=140 60 600 520,width=4.3cm, clip]{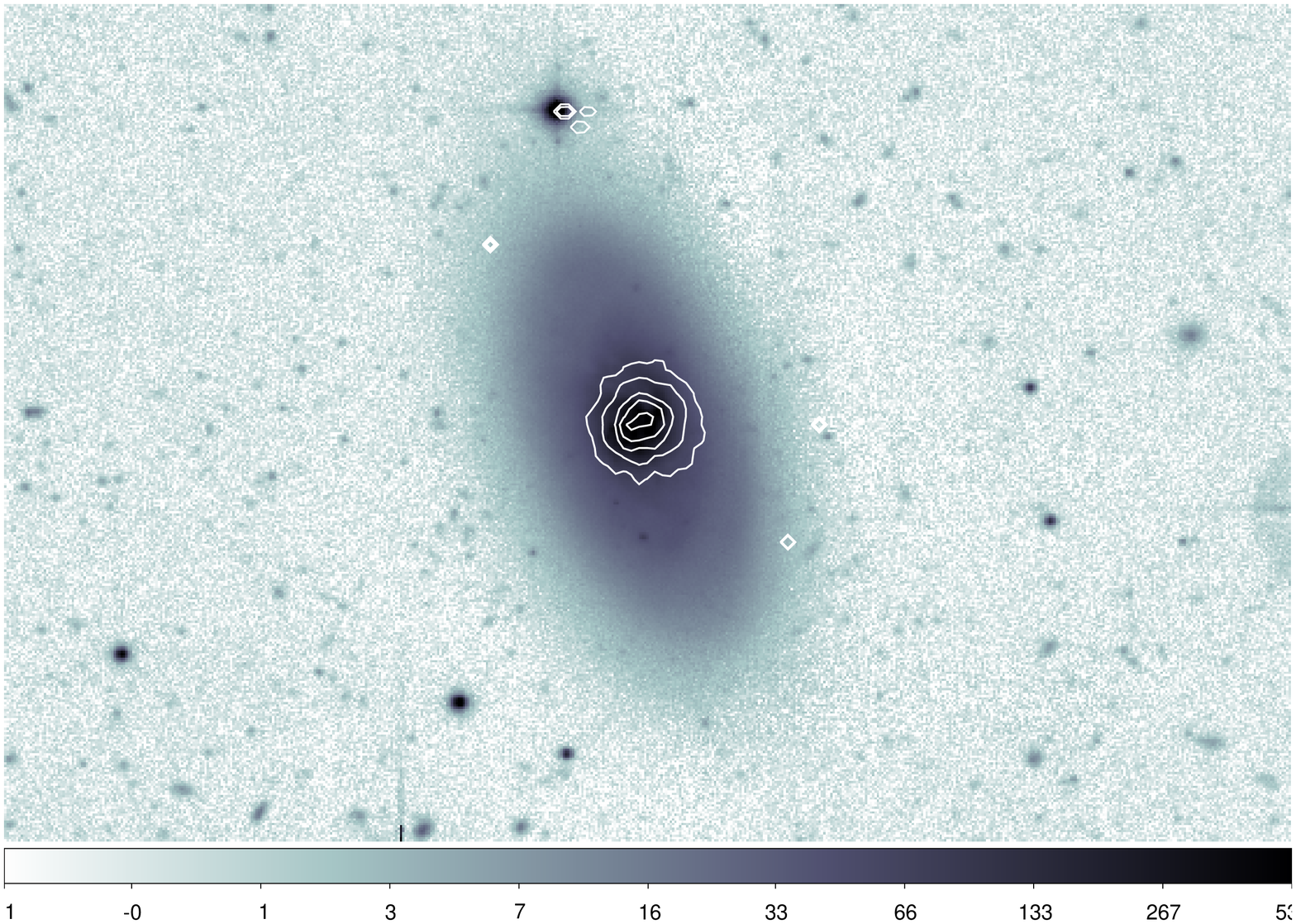}
\includegraphics[bb=140 60 600 520,width=4.3cm, clip]{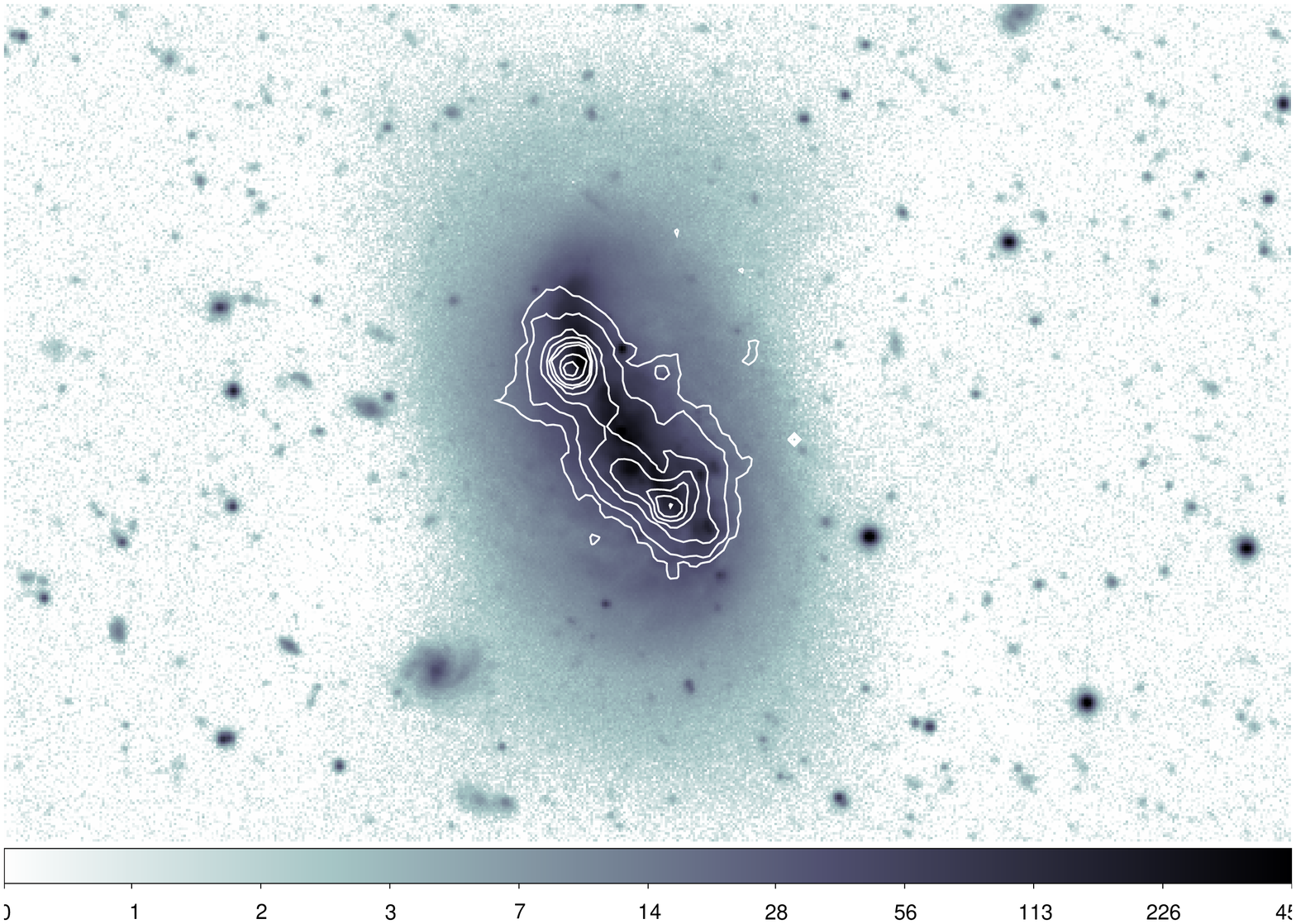}
\includegraphics[bb=140 60 600 520,width=4.3cm, clip]{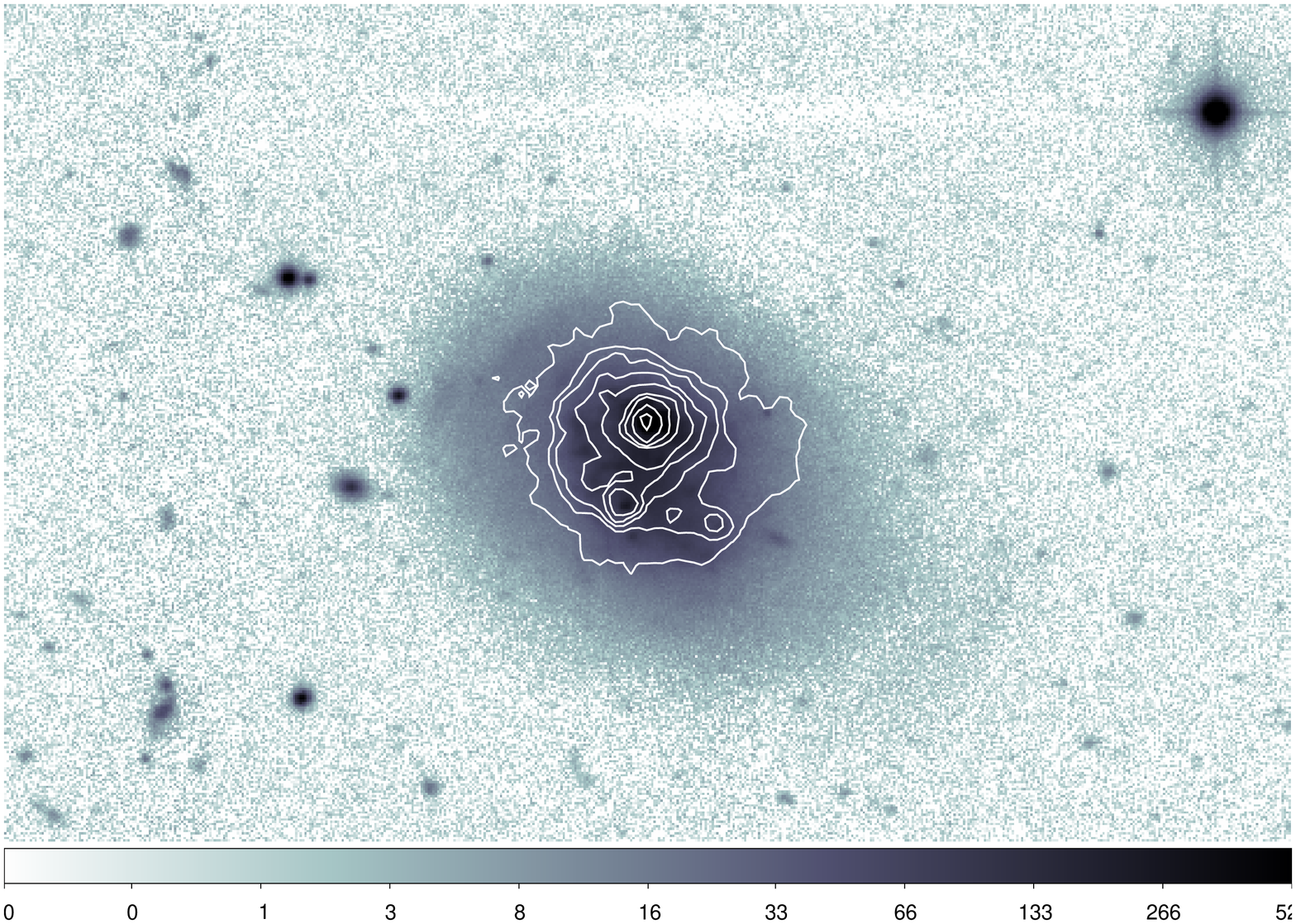}
\includegraphics[bb=0 0 565 430,width=4.6cm, clip]{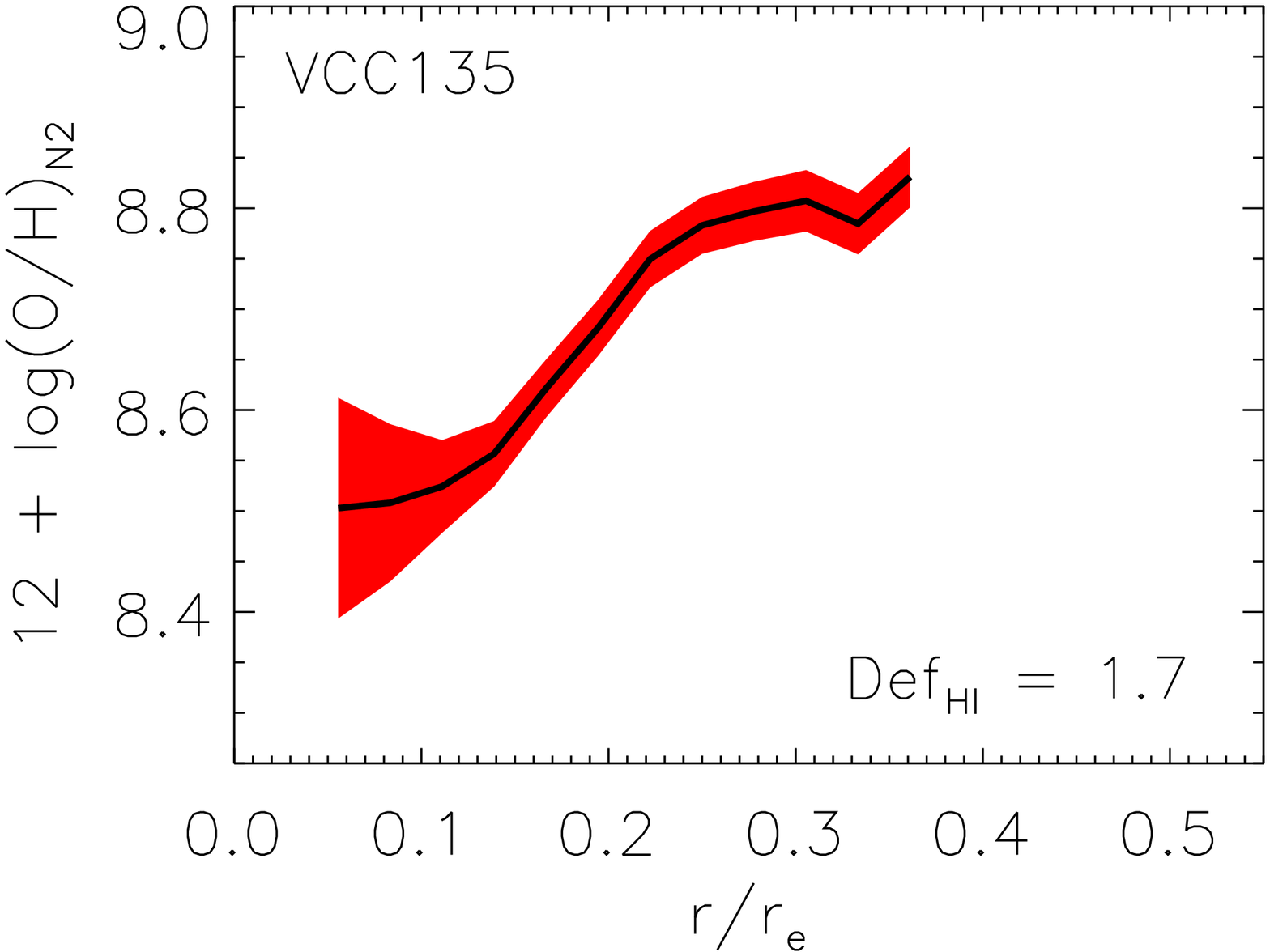}
\includegraphics[bb=40 0 565 430,width=4.3cm, clip]{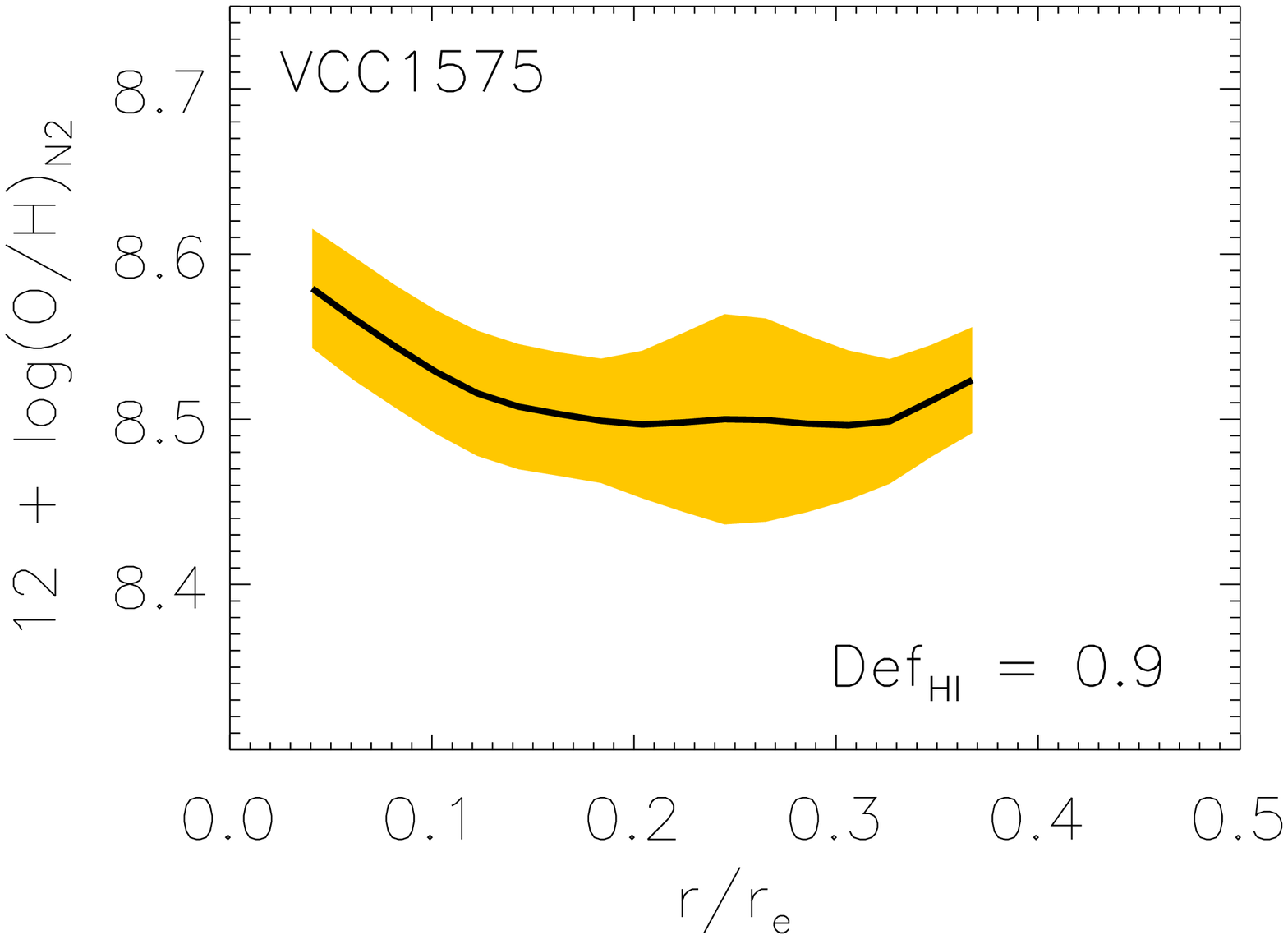}
\includegraphics[bb=40 0 565 430,width=4.3cm, clip]{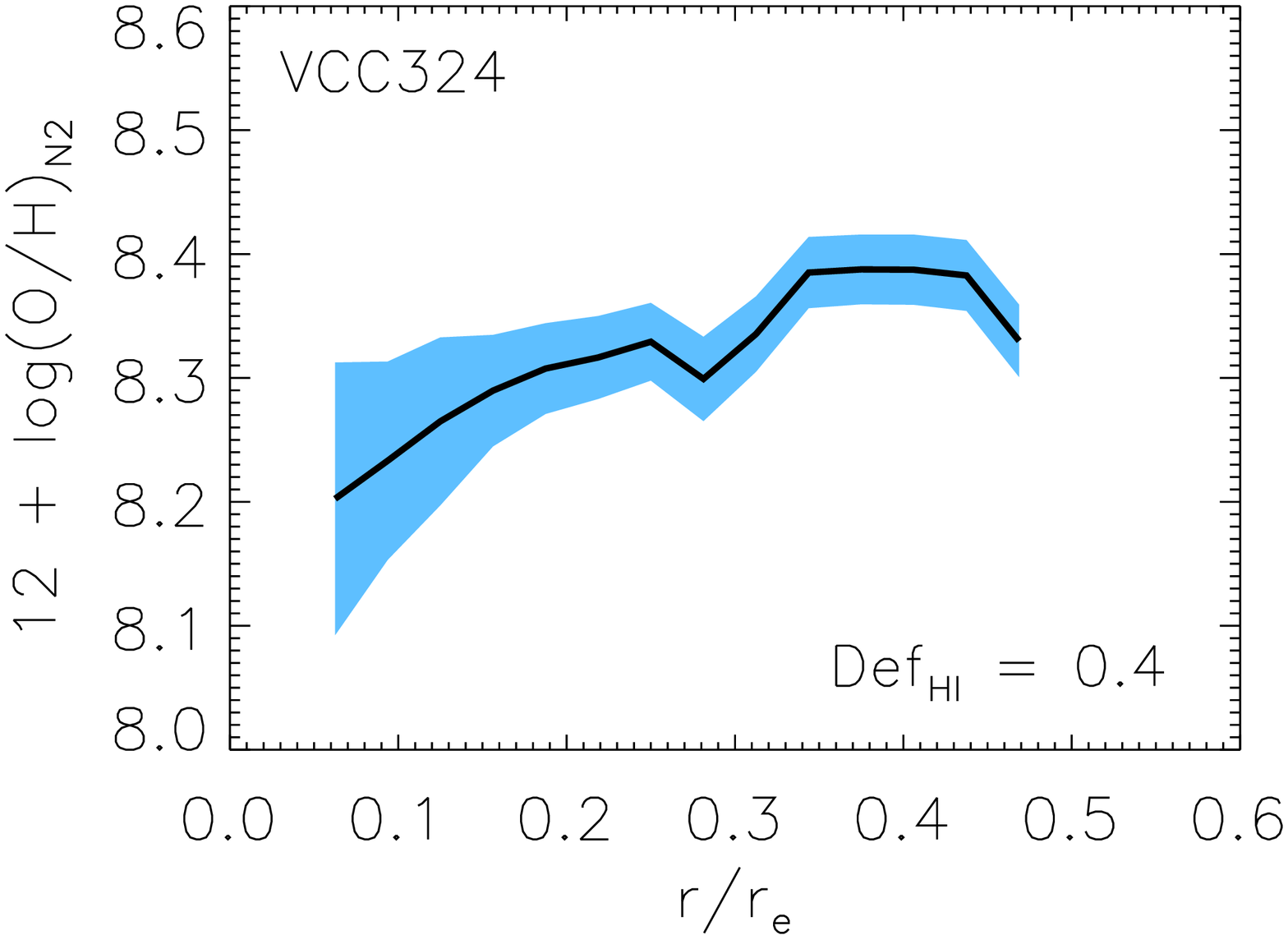}
\caption{Top panels: H$\alpha$ emission contours of three Virgo SFDGs overlaid on the optical images ($g$-band). 
The galaxies have three different levels of \hi\ deficiency. VCC~135 (to the left) is the most \hi-deficient of the sample,
while VCC~324 (to the right) is moderately \hi-deficient. The H$\alpha$ emission is more centrally concentrated in the galaxies 
with a lower \hi\
content.
Bottom panels: Oxygen abundance radial profiles of  the three galaxies. VCC~135 and VCC~324 show a peculiar inverted metallicity
gradient. 
}
\label{fig5}
\end{center}
\end{figure}

We recently studied a population of SFDGs in Virgo and addressed their dust, atomic,
and molecular gas properties (Grossi et al. 2015, Grossi et al. 2016). We found that the interaction
with a cluster environment is removing their atomic gas and dust components, as suggested by the decrease of the dust
-to-stellar mass ratio with increasing \hi\ deficiency (Fig. \ref{fig4}, left panel).
Regardings the molecular gas component, because of the small number statistics of CO detections at high \hi-deficiencies 
(empty circles in the right panel of Fig. \ref{fig4}), 
it is difficult to draw strong conclusions. Nonetheless, comparing our sample and the late-type galaxies in the Herschel 
Reference Survey (Boselli et al. 2014) it appears that the molecular gas is less affected than dust and atomic gas. 
Moreover, it results that \hdue\ becomes the dominant gaseous component 
in \hi-deficient galaxies (Grossi et al. 2016, Mok et al. 2017). 

The ionised phase is also a good tracer of stripped gas
in dense regions: late-type galaxies in nearby clusters  show
extended ($\sim$ 50 kpc) tails of ionised gas (Boselli \&
Gavazzi 2014) or truncated H$\alpha$ discs (Koopman \& Kenney 2004). 
We are carrying out a pilot survey with integral field unit spectroscopy observations of a
subset of far-infrared-detected dwarfs with different levels of \hi\ deficiencies (i.e. from none to high, 0.1
$<$ $Def_{HI}$ $<$ 1.7) using the PMAS/PPAK instrument at the 3.5m Calar Alto telescope (Grossi et al., in prep.).
The dwarfs are
at different stages of interaction with the cluster environment, thus they can give different snapshots
of the evolutionary path from star-forming to more passive systems. 
Our observations clearly show that the extension of the H$\alpha$ emission 
in galaxies with higher \hi-deficiency is truncated compared to those with a normal \hi\ content (Fig. \ref{fig5}, top panels). 
Interestingly, two galaxies in the sample show an inverted metallicity gradient
(Fig. \ref{fig5}, bottom panels) which are usually observed
in isolated dwarf galaxies (see \S 2.1) or at high redshift (Cresci et al. 2010).

Ram-pressure stripping can efficiently remove the ISM of low-mass star-forming systems on very short time
scales ($\sim$ few hundreds Myrs), and it can  transform gas-rich star-forming systems into
gas-poor quiescent objects on time scales of the order
of 0.8 - 1.7 Gyr (Boselli \& Gavazzi 2006; Boselli et al. 2014).
However, 
ram pressure stripping is not the only mechanism responsible for producing dEs in clusters and gravitational interactions must 
also play a role (Mastropietro et al. 2005, Mistani et al. 2016).

Structurally, dEs are more complex than they might seem. Beneath their regular and smooth appearance some dEs 
present late-type features such as discs, spiral arms, 
or blue centers (Jerjen et al. 2000; Lisker et al. 2006, 2007, 2009;  
Janz et al. 2012), and in some cases they can host an atomic gas or dust component 
(di Serego Alighieri 2007, 2013; Hallenbeck et al. 2017).
They also have a wide range of kinematic properties, including clear signs of rotation 
(Pedraz et al. 2002, Toloba et al. 2009),
with rotation curves comparable to those of late-type spiral galaxies of similar luminosity (Toloba et al. 2011).
Rotationally supported dEs in Virgo are mainly located at the periphery of the cluster and they host younger stellar
populations, while pressure-supported systems have predominantly-old stellar populations and they are found in the
inner regions (Toloba et al. 2009).
These properties suggest a formation scenario where
more than one process may determine their final evolution.
The transformation from a star-forming to a passive dwarf
could involve an initial phase of ram-pressure stripping as the galaxy enters a dense environment for the first time,
where most of its ISM is removed 
(Conselice et al. 2003; Boselli \& Gavazzi 2006; Boselli et al. 2014), followed by tidal interactions that kinematically heat the stellar disc 
modifying the galaxy
morphology (Gnedin 2003; Lisker et al. 2009; Toloba et al. 2015).

\begin{figure}[t]
\begin{center}
\includegraphics[bb=30 30 565 585,width=5.5cm]{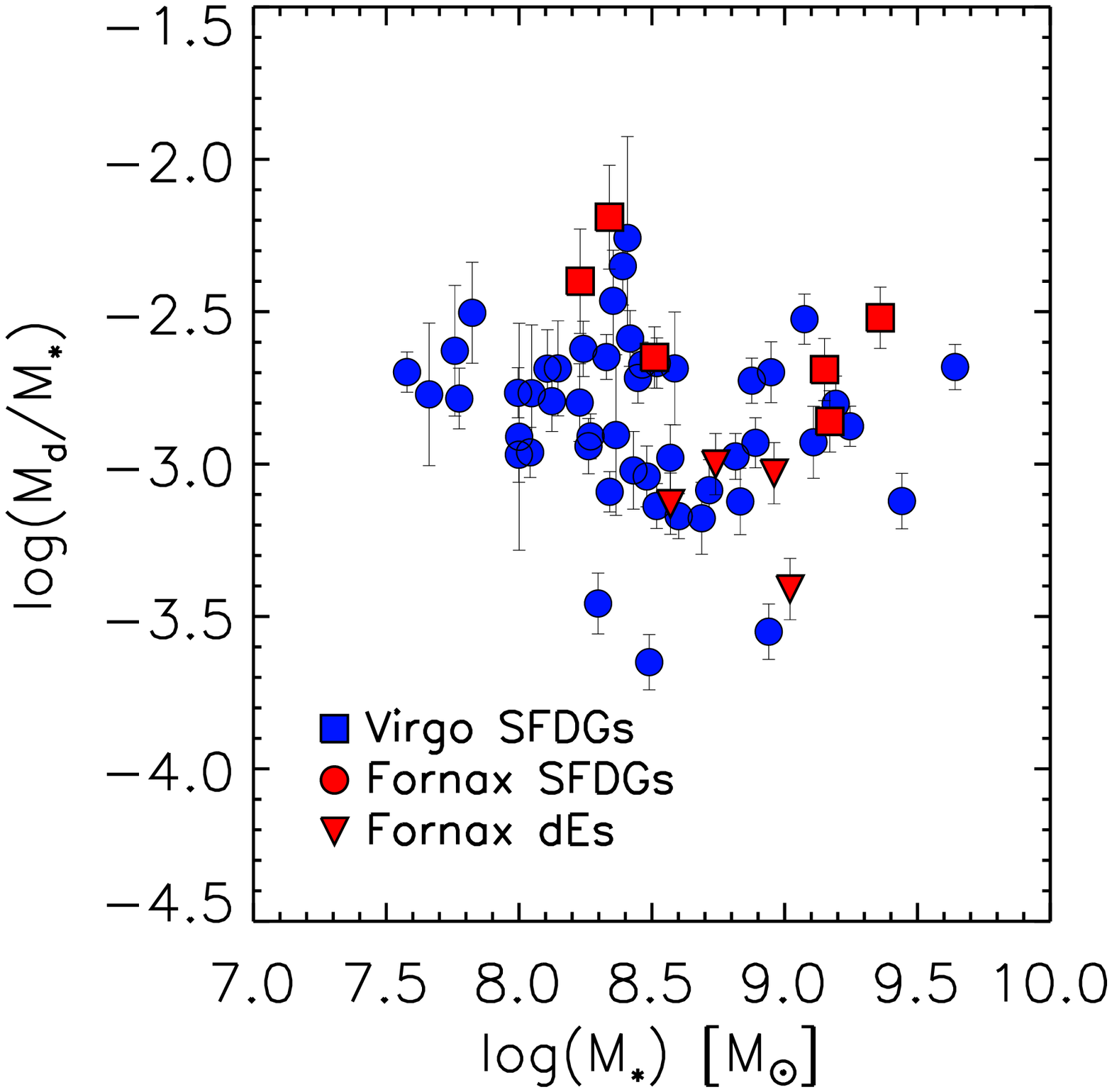}
\includegraphics[bb=50 -10 700 460,width=7.3cm,clip]{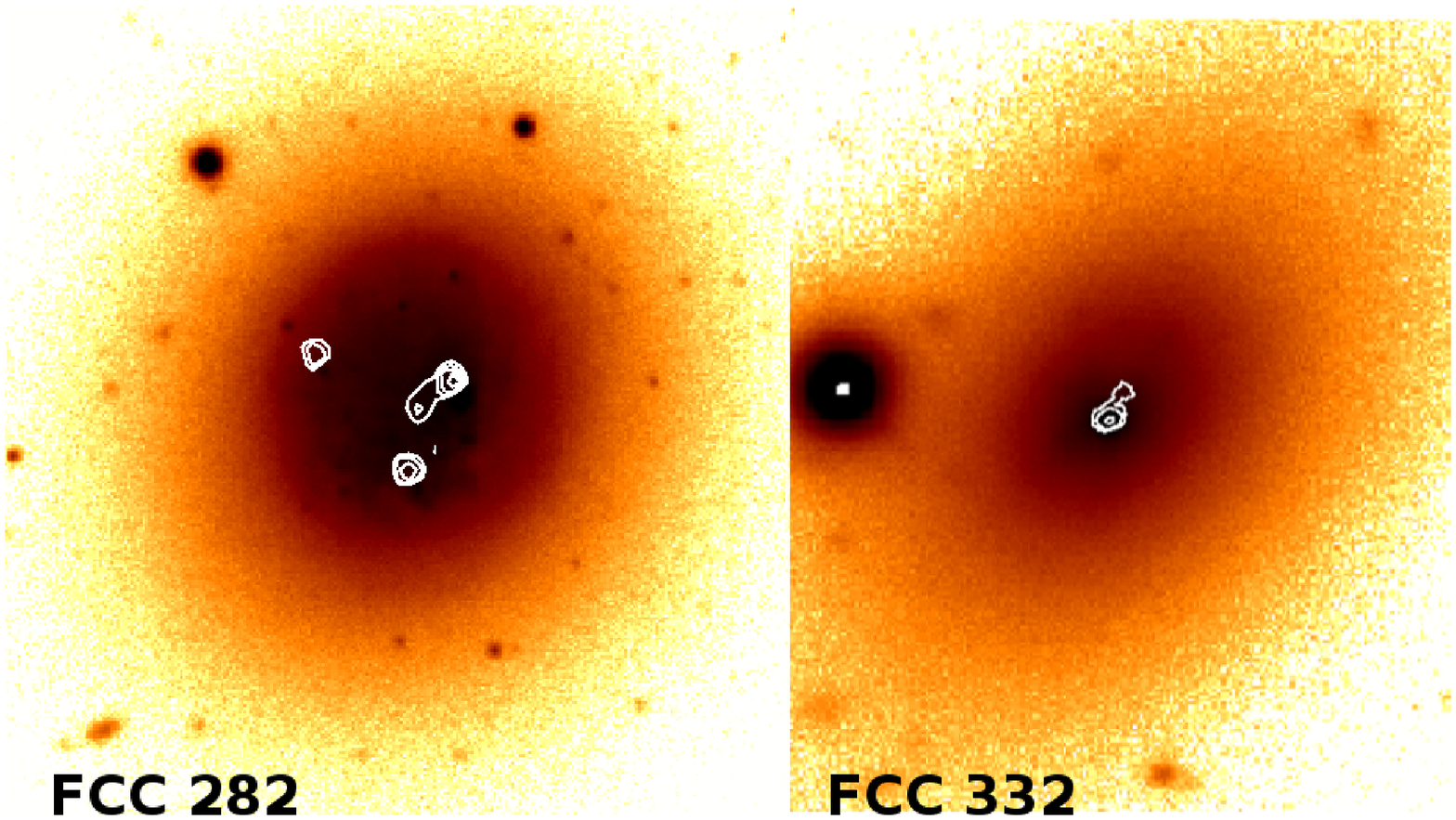}
\caption{Left panel: Comparison between the dust-to-mass ratio of Virgo and Fornax dwarfs detected in the $Herschel$ surveys of the clusters. 
Central and right panels: Example of two Fornax dEs (FCC~282, FCC~332) showing residual star formation activity.
H$\alpha$ + [N{\sc ii}] emission (contours) overlaid on the $r$-band image obtained with Gemini/GMOS-S.}
\label{fig6}
\end{center}
\end{figure}

\subsection{The Fornax cluster}

It is important to study other nearby galaxy clusters and verify if these trends apply also to 
dense environments with different characteristics than Virgo.
Fornax is the second largest galaxy cluster within a distance of 20 Mpc (Blakeslee et al. 2009).
The main structure is very compact (r$_{vir}$ $\simeq$ 750 kpc, Drinkwater et al. 2001) and
consists of 22 galaxies brighter than $M_B < -18$ mag and around
200 fainter galaxies (Ferguson 1989).
Fornax is a factor of seven less massive than Virgo (Jordan et al. 2007), 
thus 
it bridges the gap between evolved groups and more massive clusters.
It has a more regular shape and it is more dynamically
evolved than Virgo, as indicated by 
 the high early-type galaxy fraction within the virial radius (87\% including massive and dwarf ETGs; Ferguson
1989). 
However, it shows evidence of a substructure 
at about 3$^{\circ}$ southwest from the cluster centre, characterized by 
a predominant late-type population of galaxies that are infalling towards the main structure
(Drinkwater et al. 2001). 
Previous \hi\ surveys found 
that Fornax galaxies present a moderate \hi\ deficiency,
and that there is a considerable deficit of \hi-rich galaxies in the centre of the cluster (Schroeder et al. 2001, Waugh et al. 2002).

Fornax has not been studied as well as Virgo yet, however different multi-wavelength surveys have been recently completed or 
are currently ongoing.
The central region of the cluster out to the virial radius 
has been observed with $Herschel$ (Davies et al. 2013, Fuller et al.
2014); members with a stellar mass $\gtrsim 10^ 8$ \msun\ have been targeted with ALMA to search for 
CO(1-0) emission (Zabel et al. 2018);
a region of 12 square degrees that covers the main cluster and the infalling substructure centred on NGC~1316
will be studied at 21-cm with the MEERKAT telescope (Serra et al. 2016); two deep optical surveys  
of approximately 30 square degrees around the cluster
 have been recently completed, the Fornax Deep Survey (FDS, Venhola et al. 2018) and the Next Generation Fornax Cluster Survey (NGFCS,
Ordones-Briceno 2018).

The population of dwarf galaxies in Fornax is dominated by early-type dwarfs as in Virgo 
(Ferguson et al. 1989), and
the late-to-early-type ratio increases with radius (de Rijcke et al. 2010). 
The larger velocity dispersion of Fornax SFDGs and their more extended
spatial distribution compared to the giant members indicate that
they are an infalling population recently accreted from the cluster (Drinkwater et al. 2001).
The fraction of Fornax SFDGs detected by the $Herschel$ within the virial radius is lower
compared to Virgo (31\% against 47\%, Fuller et al. 2014), but both populations appear to have similar dust properties 
(Fig. \ref{fig6}, left panel). 
On the other hand, Fornax early-type dwarfs present a higher rate of star formation than expected by their morphological 
classification; 30\% of the dE population studied by Drinkwater et al. (2001)
present a significant H$\alpha$ emission (EW $>$ 3 \AA; Fig. \ref{fig6}). 
Some of these dEs also host a dust (Fuller et al. 2014) 
and a molecular gas component (Zebel et al. 2018).
Thus they could represent a link between more vigorously star-forming dwarfs and
quenched dEs in Fornax, and the ongoing surveys of the cluster will allow to shed light on the
evolutionary history of these systems.

\begin{discussion}

\discuss{Q1}{For star-forming dwarfs in voids, which process is more responsible for the low oxygen abundances in the ionised gas, 
inflow from the IGM low Z gas or that there hasn't been many galaxy-galaxy mergers (i.e. generations of massive stars)?}

\discuss{A1}{It is difficult to answer which process is dominating without a systematic study of either the \hi\ morphology 
or the radial metallicity variation in void dwarf galaxies, which could provide evidence of ongoing gas accretion. So far, it seems that 
both mechanisms can be responsible for the observed low metallicities.}

\discuss{Q2}{Speaking of \hi\ deficiencies requires to normalize the \hi\ content to "normal" galaxies. How is this feasible for the inhomogeneous samples of dIrrs?
}

\discuss{A2}{Blind \hi\ surveys like ALFALFA are providing large samples of field late-type dwarfs that can be used as 
reference to assess the \hi\ content of normal unperturbed dwarf systems. The \hi-deficiency parameter for late-type dwarf galaxies has
been recently recalibrated by Gavazzi et al. (2013).}

\discuss{Q3}{Given that very few dwarfs in Fornax have been detected with $Herschel$, 
can you tell us whether the optical image scan give us some clues about when a galaxy contains dust?
}

\discuss{A3}{This kind of analysis can be performed with the deep multi-imaging optical surveys of Fornax (i.e. the FDS and the NGFCS). 
We don't have access to these data sets, thus we cannot say
much about the dust content of galaxies not detected with $Herschel$.   
}

\end{discussion}

\end{document}